\documentclass[conference]{IEEEtran}
\IEEEoverridecommandlockouts

\usepackage{cite}
\usepackage{amsmath,amssymb,amsfonts}
\usepackage{algorithmic}
\usepackage{graphicx}
\usepackage{textcomp}
\usepackage{xcolor}
\usepackage[euler]{textgreek}
\usepackage{tabularx}
\usepackage{comment}
\usepackage{makecell}
\usepackage{caption}
\usepackage{subcaption}
\usepackage{dsfont}
\usepackage{amsthm}
\theoremstyle{plain}

\theoremstyle{remark}

\def\BibTeX{{\rm B\kern-.05em{\sc i\kern-.025em b}\kern-.08em
    T\kern-.1667em\lower.7ex\hbox{E}\kern-.125emX}}

\usepackage{fancyhdr}
\fancypagestyle{preprint}{
  \fancyhf{}
  
  \fancyhead[C]{\fontsize{9}{10}\selectfont \textcolor{gray}{This work has been accepted for publication in \textbf{IEEE Globecom 2025}. \\ Please cite the official IEEE version once available.}}
}

\usepackage[hidelinks]{hyperref}
\hypersetup{
    pdftitle={Adaptive Underwater Acoustic Communications with Limited Feedback: An AoI-Aware Hierarchical Bandit Approach},
    pdfauthor={Fabio Busacca, Andrea Panebianco, Yin Sun},
    pdfsubject={Accepted for publication in IEEE Globecom 2025},
    pdfkeywords={Underwater Communications, Adaptive Modulation, Power Control, Reinforcement Learning, Multi-Armed Bandit, Age of Information}
}
    
\begin{document}

\title{Adaptive Underwater Acoustic Communications with Limited Feedback: An AoI-Aware Hierarchical Bandit Approach
\thanks{This work was supported in part by the National Science Foundation (NSF) under Grant CNS-2239677, and in part by the European Union under the Italian National Recovery and Resilience Plan (NRRP) of NextGenerationEU, within the “Telecommunications of the Future” partnership (PE0000001 – program “RESTART”). The authors are listed alphabetically.

\textbf{Accepted for publication in IEEE Globecom 2025. © 2025 IEEE. Personal use of this material is permitted. Permission from IEEE must be obtained for all other uses. Please cite the official version when available.}}}

\author{
\IEEEauthorblockN{Fabio Busacca$^{o}$, Andrea Panebianco$^*$$^{\S}$, Yin Sun$^{\S}$}
{ $^o$\textit{University of Catania, Italy}}
{$^*$\textit{University of Palermo, Italy}}
{$^{\S}$\textit{Auburn University, Alabama, USA}}
}
\maketitle
\thispagestyle{preprint}

\begin{abstract}
Underwater Acoustic (UWA) networks are vital for remote sensing and ocean exploration but face inherent challenges such as limited bandwidth, long propagation delays, and highly dynamic channels. These constraints hinder real-time communication and degrade overall system performance. To address these challenges, this paper proposes a bilevel Multi-Armed Bandit (MAB) framework. At the fast inner level, a Contextual Delayed MAB (CD-MAB) jointly optimizes adaptive modulation and transmission power based on both channel state feedback and its Age of Information (AoI), thereby maximizing throughput. At the slower outer level, a Feedback Scheduling MAB dynamically adjusts the channel-state feedback interval according to throughput dynamics: stable throughput allows longer update intervals, while throughput drops trigger more frequent updates. This adaptive mechanism reduces feedback overhead and enhances responsiveness to varying network conditions. The proposed bilevel framework is computationally efficient and well-suited to resource-constrained UWA networks. Simulation results using the DESERT Underwater Network Simulator demonstrate throughput gains of up to 20.61\% and energy savings of up to 36.60\% compared with Deep Reinforcement Learning (DRL) baselines reported in the existing literature.
\end{abstract}


\section{Introduction} \label{sec:intro}

UnderWater (UW) networks are attracting growing attention from academia and industry, enabled by advances in acoustic communication technologies \cite{akyildiz2005underwater,murad2015survey}. They support real-time data exchange in remote, harsh environments for applications such as environmental monitoring, exploration, and disaster response. However, UnderWater Acoustic (UWA) networks face intrinsic constraints—limited bandwidth, long propagation delays, and highly dynamic channels—that require adaptive protocols with low computational and signaling overhead. In particular, frequent feedback over costly acoustic channels can overwhelm the network, making intelligent scheduling critical to sustain throughput while limiting signaling and energy consumption.

A promising solution is Adaptive Modulation (AM), which dynamically selects Modulation Schemes (MSs) based on real-time channel conditions \cite{busacca2024survey}. Adjusting transmission power ($P$) can further improve reliability. Yet, adapting these parameters in highly dynamic UW environments remains difficult, requiring fast responsiveness within the limited computational resources of UW devices.

Model-based approaches have long been applied to improve UW network performance \cite{Sadhu2023jscc,busacca2024comparative}, predicting channel dynamics to guide decisions such as MS selection or data scheduling. Their effectiveness, however, is limited by the non-linear and unpredictable nature of UW environments, where factors like temperature, salinity, and water motion introduce uncertainties that are difficult to model.

To address these limitations, Deep Reinforcement Learning (DRL) has been explored for dynamic adaptation in UW environments. Although DRL provides accuracy and adaptability, it incurs high computational cost, slow convergence, and extensive training time, limiting its use in dynamic, resource-constrained UWA networks \cite{zhang2021reinforcement,Wang2021rl}. In contrast, lightweight Multi-Armed Bandit (MAB) algorithms have recently attracted attention for their efficiency and ability to adapt to changing conditions without large datasets or heavy training \cite{busacca2025amuse}. They offer a computationally efficient alternative that balances performance and adaptability, making them well-suited for real-time UW applications.

This paper presents a bilevel MAB framework that jointly optimizes throughput and feedback overhead through adaptive transmission control and feedback scheduling. The main contributions are as follows:

\begin{itemize}
    \item We propose a novel fully distributed bilevel MAB framework for UWA networks. The inner layer employs a Contextual Delayed MAB (CD-MAB) to jointly adapt modulation and power, maximizing throughput and reliability. The outer layer uses a Feedback Scheduling MAB to tune the feedback interval, extending it under stable performance and shortening it under degradation, thereby controlling context freshness, shaping the Age of Information (AoI), and reducing overhead. To the best of our knowledge, this is the first distributed bilevel MAB framework integrating adaptive modulation, power control, and AoI-aware feedback scheduling for UWA networks.
    
\begin{figure} [!ht]
\centering 
\includegraphics[width=\linewidth]{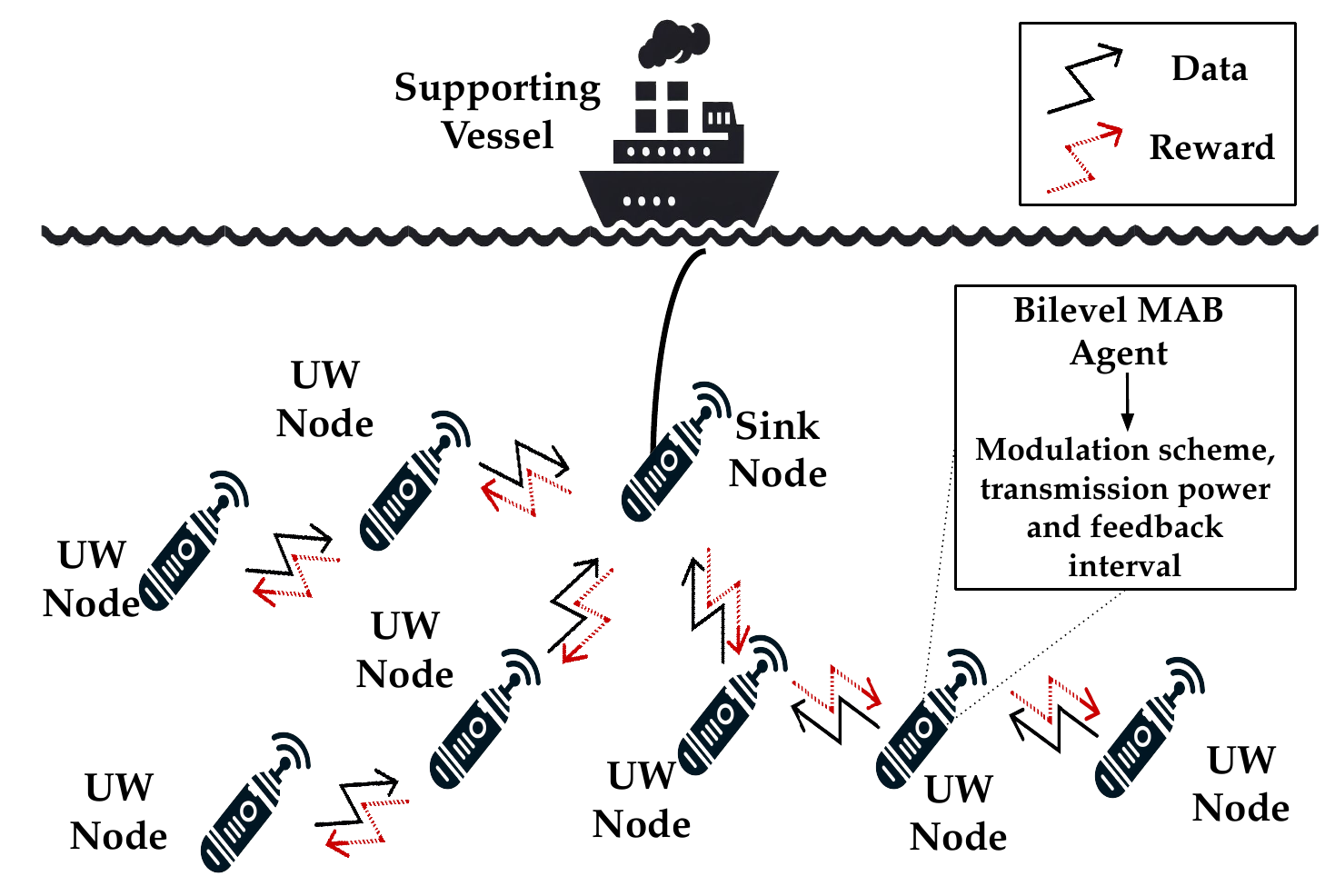} 
\caption{Underwater Acoustic Network architecture.} 
\label{sysarch} 
\end{figure}
    
    \item We validate the proposed framework through extensive simulations using the DEsign, Simulate, Emulate and Realize Test-beds (DESERT) UW simulator \cite{DESERT}, incorporating realistic environmental conditions and dynamic channel variations for a comprehensive assessment.
    \item We compare our framework with existing DRL-based schemes and show that it \textbf{achieves up to 20.61\% higher throughput and 36.60\% energy savings}, thanks to distributed design, joint modulation and power adaptation, and dynamic feedback scheduling. Unlike prior centralized DRL solutions \cite{mashhadi2021deep,8703432} with fixed MS, $P$, and feedback frequency, our benchmarks are drawn from terrestrial wireless networks, given the limited availability of effective RL methods specifically tailored to UWA environments.
\end{itemize}

\section{Reference System} \label{sysarc}

This section presents the reference scenario used to evaluate the proposed UWA network, modeled with DESERT.

The proposed architecture, shown in Fig.~\ref{sysarch}, includes a set of $\mathcal{U}$ Internet of Underwater Things (IoUT) nodes deployed in a 3D shallow-water environment. Each node $u \in \mathcal{U}$ communicates with a central \textit{Sink} via single- or multi-hop acoustic routing. The Sink aggregates data from the nodes and forwards it via a wired link to a Supporting Vessel, which provides power and connectivity to the Sink. All nodes, except \textcolor{black}{for} leaf nodes, act as both transmitters (Txs) and receivers (Rxs) \textcolor{black}{in a half-duplex fashion}, with Tx–Rx interactions occurring at the link level, i.e., between directly connected neighbors.

\begin{figure} [!h] 
\centering 
\includegraphics[width=0.8\linewidth]{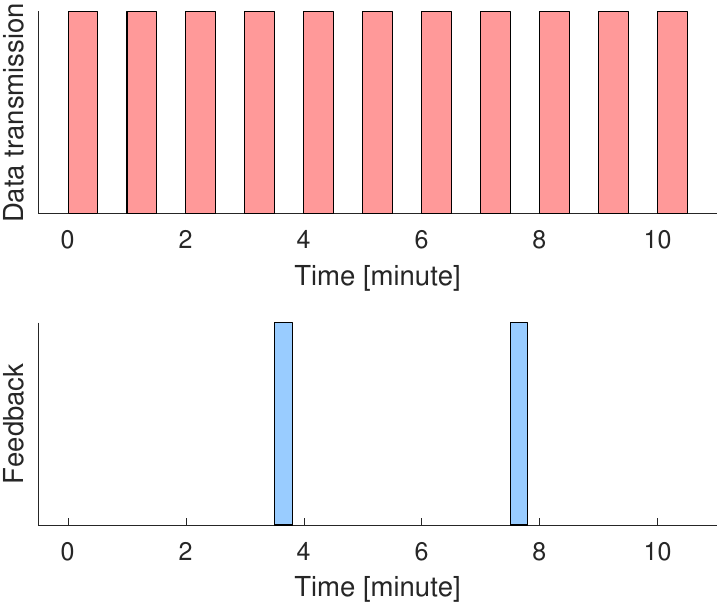} 
\caption{\textcolor{black}{Data is transmitted periodically, while feedback is not sent after every event. Our algorithm adapts the number of consecutive transmissions before the Rx sends feedback.}}
\vspace{-0.45cm}
\label{fig:transmission_time} 
\end{figure}

The Tx periodically transmits sensor data packets embedding the selected $P$ level, allowing the Rx to estimate the Signal-to-Noise Ratio (SNR) based solely on channel conditions, independent of the power adaptation policy. The Rx computes SNR and throughput from the received signal and periodically returns this information via dedicated single-hop feedback links.
Feedback is sent only at the end of each feedback interval, summarizing performance and reporting the estimated SNR from the most recent channel observation. This aggregated approach reduces signaling overhead while still providing the Tx with sufficient information to update its transmission strategy.

At the end of each interval, the Tx requests feedback from the Rx to update its strategy and remain synchronized. Upon receiving delayed feedback, the Tx node, which runs both the CD-MAB and the Feedback Scheduling MAB, updates its decisions. The CD-MAB selects the optimal \textcolor{black}{Modulation Scheme (MS)} (among BPSK, 8-PSK, and 16-PSK) and \textcolor{black}{transmission power $P$ (discretized transmission power, with three levels: low, medium, and high)} to maximize throughput, \textcolor{black}{while the Feedback Scheduling MAB selects the feedback interval from $\{4,7,10\}$ minutes, and hence balances throughput and feedback cost.} More frequent feedback improves network tracking, indirectly reducing AoI, while longer intervals reduce signaling overhead at the cost of less timely channel information. This process is illustrated in Fig.~\ref{fig:transmission_time}.
In realistic shallow-water scenarios, SNR typically ranges from 10 dB in poor conditions to 40 dB in optimal ones \cite{zhang2009time,yang2012properties}. To capture this variability with low overhead, we quantize the SNR into three non-uniform intervals, reflecting its nonlinear impact on communication performance. At low SNR (10–18 dB), even small variations significantly affect Bit Error Rate (BER) and throughput, requiring finer granularity. In the medium range (18–30 dB), performance improves more gradually, while at high SNR (above 30 dB), it saturates, making finer distinctions unnecessary.
\begin{itemize}
    \item \textbf{Low Quality} [$10$, $18$] dB: narrow range due to BER and packet loss being highly sensitive to minor SNR changes.
    \item \textbf{Medium Quality} ($18$, $30$] dB: broader interval reflecting more gradual performance improvements.
    \item \textbf{High Quality} ($30$, $40$] dB: performance saturates, further increases provide minimal additional gains.
\end{itemize}

\section{Age of Information} \label{aoi}

In the proposed framework, the Age of Information (AoI) quantifies feedback freshness at the Tx. It measures the time elapsed since the last feedback packet, reflecting how outdated the channel knowledge of the Tx is \cite{yates2021age,shisher2024monotonicity}.

Let $k \in \mathbb{N}$ denote the index of feedback epochs, with $t_k$ the time of the $k$-th feedback and $Q_k = t_k - t_{k-1}$ the interval duration. During $Q_k$, the AoI increases \textcolor{black}{discretely by one unit per slot}, and resets to zero at $t_k$. For any $t \in [t_{k-1}, t_k)$, the AoI evolves as:

\begin{equation}
\Delta(t) = t - t_{k-1}.
\end{equation}
This time-evolving metric quantifies channel information staleness,  guiding adaptive feedback decisions.

\section{Our Hierarchical Bilevel MAB Approach}

\textcolor{black}{To jointly optimize throughput and feedback overhead, we formulate a bilevel MAB framework solving a unified problem. The inner Contextual Delayed Multi-Armed Bandit (CD-MAB) adapts modulation and power based on feedback, while the outer MAB adjusts the feedback interval to balance signaling cost and context freshness. Together, they coordinate to maximize long-term performance in dynamic UW settings.}

\subsection{Contextual Delayed Multi-Armed Bandit (Inner Loop)} \label{CMAB}

The Tx uses delayed contextual feedback and an Upper Confidence Bound (UCB)-based policy~\cite{ucb1,chen2024contextual}.  Within each feedback interval $[t_{k-1}, t_k)$, \textcolor{black}{discretized into time slots $t$ (each lasting 1 minute in wall-clock time)}, the CD-MAB selects an action $a_t \in \mathcal{A}$, where $\mathcal{A}$ is the set of feasible MS–$P$ pairs. The context ($X_t$) includes the most recent SNR estimate ($\hat{\eta}$) and its associated AoI, indicating channel information freshness, so $X_t = (\hat{\eta}_{t-\Delta(t)}, \Delta(t))$.

Since the reward is observed only at the end of each feedback interval, the reward $r_k$ received at $t_k$ is the sum of throughput values from all actions taken during $[t_{k-1}, t_k)$. This design avoids slot-wise rewards, ensuring constant, minimal overhead. To assign credit to individual actions, we apply a Uniform Credit Assignment (UCA) strategy, distributing $r_k$ equally over the interval. Each action $a_t$, executed under context $X_t$, is thus assigned a per-action reward:

\begin{equation}
g_t = \frac{r_k}{|\mathcal{T}_k|}, \quad \text{for } t \in [t_{k-1}, t_k),
\end{equation}
where $|\mathcal{T}_k|$ is the number of actions taken during the interval.

The CD-MAB learns an optimal transmission policy $\pi_{\text{tx}}$ that maps each context $X_t$ to an action $a_t$, \textcolor{black}{maximizing the expected throughput over a finite time horizon $\mathcal{T}$ (i.e., the total number of time slots observed across all intervals)}:

\begin{equation} 
\max_{\pi_{\text{tx}}} \quad \mathbb{E} \left[ \sum_{t=1}^{\mathcal{T}} g_t \right], \quad \text{with } a_t = \pi_{\text{tx}}(X_t).
\end{equation}

The action selection follows the UCB criterion:

\begin{equation}
a_t = \arg\max_{a \in \mathcal{A}} \left( \hat{\mu}_t(a, X_t) + \sqrt{\frac{c\log n_t}{N_t(a, X_t)}} \right),
\end{equation}
where $\hat{\mu}_t(a, X_t)$ is the empirical mean reward for action $a$ under context $X_t$, $c$ is the exploration–exploitation parameter, $n_t$ is the number of total decisions, and $N_t(a, X_t)$ is the number of times $a$ was selected under that context.

Action value estimates are updated in two stages. First, an initial update is made upon action selection, assuming immediate feedback. Then, when the delayed reward $r_k$ is received at the end of the interval, the estimate is corrected using the per-action reward $g_t$ from the UCA scheme. The correction uses an incremental average:
\begin{equation}
\hat{\mu}_t(a_t, X_t) \leftarrow \hat{\mu}_t(a_t, X_t) + \frac{1}{N_t(a_t, X_t)} \left( g_t - \hat{\mu}_t(a_t, X_t) \right),
\end{equation}
for all $t \in [t_{k-1}, t_k)$, \textcolor{black}{where $N_t(a_t, X_t)$ denotes the cumulative number of times the pair $(a_t, X_t)$ has been selected up to time $t$}. This allows responsiveness despite delay and accurate long-term learning.

\subsection{Feedback Scheduling MAB (Outer Loop)} \label{feedbackMAB}

In parallel with the CD-MAB, a second non-contextual stochastic bandit agent selects the feedback interval \textcolor{black}{duration} $Q_k \in \mathcal{Q}$ at each $t_k$, where $\mathcal{Q}$ defines the allowable durations between feedback transmissions. This choice directly affects control signaling frequency and context freshness. The Feedback Scheduling MAB aims to maximize throughput while minimizing the cost of frequent feedback, which interrupts transmission and consumes energy. Although not directly optimized, a lower AoI naturally enhances throughput.

To capture this trade-off, the reward is the cumulative throughput $r_k$ during interval \textcolor{black}{$[t_{k-1},t_k]$ (with $Q_k=t_k-t_{k-1}$)}, penalized by the energy cost term proportional to feedback frequency.
\begin{equation}
r_{\text{fb},k} = \theta r_k - (1 - \theta) C_{\text{fb}} \frac{1}{Q_k}, \end{equation}
where $C_{\text{fb}}$ is the energy cost of a feedback packet, computed as the product of its transmission duration and transmission power $P$. The parameter $\theta \in [0, 1]$ tunes the trade-off between throughput and feedback transmission cost, and $\frac{1}{Q_k}$ reflects the average feedback rate. Since both $r_k$ and $Q_k$ vary at each round, $r_{\text{fb},k}$ is defined as a function of $k$.
Smaller $Q_k$ improves feedback freshness, boosting throughput at the cost of higher overhead. Longer intervals reduce signaling but risk outdated context. The agent learns to select shorter intervals when rapid adaptation is needed and longer ones when the channel is stable.
The optimization objective is:

\begin{equation}
\max_{\pi_{\text{fb}}} \quad \sum_{k=1}^{K} \left( \theta r_k - (1 - \theta) C_{\text{fb}} \frac{1}{Q_k} \right),
\end{equation}
where $\pi_{\text{fb}}$ is the feedback scheduling policy.

\begin{figure*}[!ht]
\centering
\begin{subfigure}{0.36\textwidth}
    \centering
    \includegraphics[width=\linewidth]{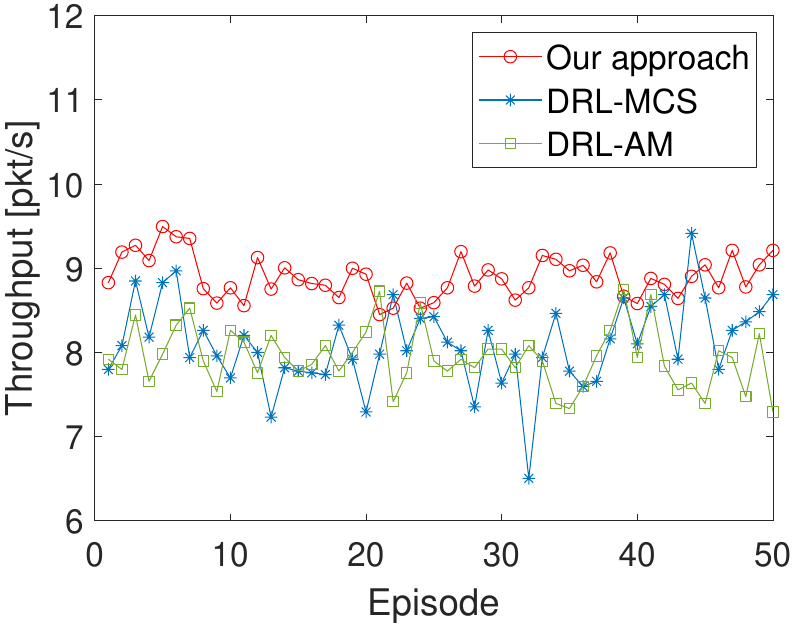}
    \caption{8-node scenario.}
    \label{fig:8nodessim}
\end{subfigure}
\hspace*{2cm}
\begin{subfigure}{0.36\textwidth}
    \centering
    \includegraphics[width=\linewidth]{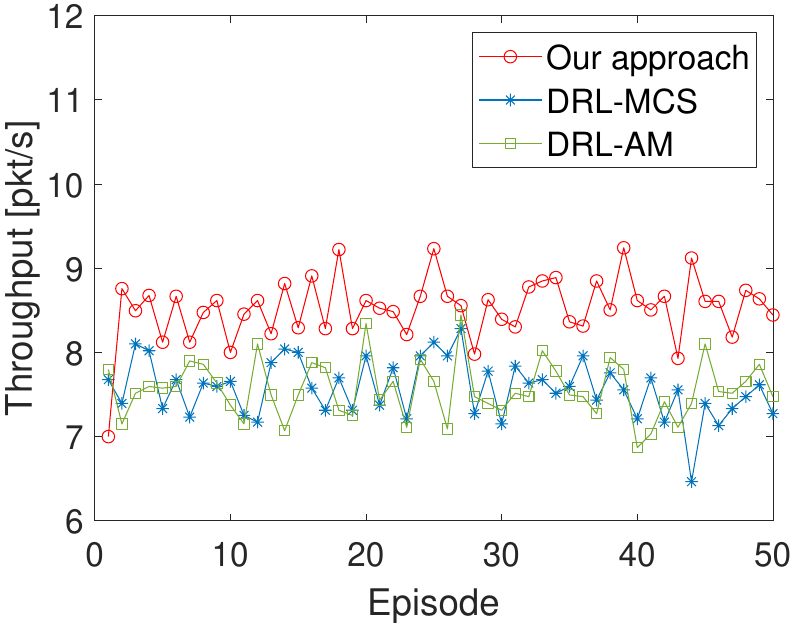}
    \caption{10-node scenario.}
    \label{fig:10nodessim}
\end{subfigure}

\caption{Throughput comparison between our approach, DRL-MCS, and DRL-AM.}
\label{fig:throughput}
\end{figure*}

\begin{figure*}[!ht]
\centering
\begin{subfigure}{0.36\textwidth}
    \centering
    \includegraphics[width=\linewidth]{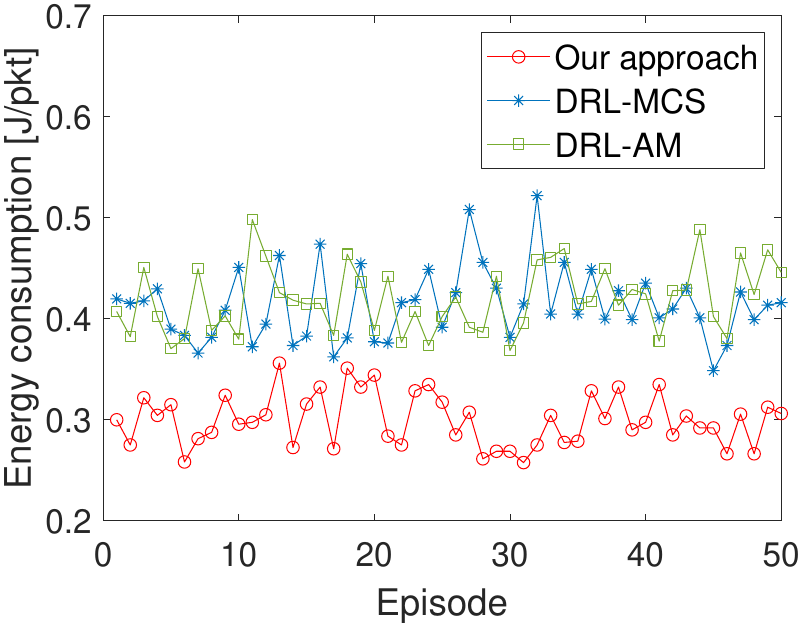}
    \caption{8-node scenario.}
    \label{fig:8nodesener}
\end{subfigure}
\hspace*{2cm}
\begin{subfigure}{0.36\textwidth}
    \centering
    \includegraphics[width=\linewidth]{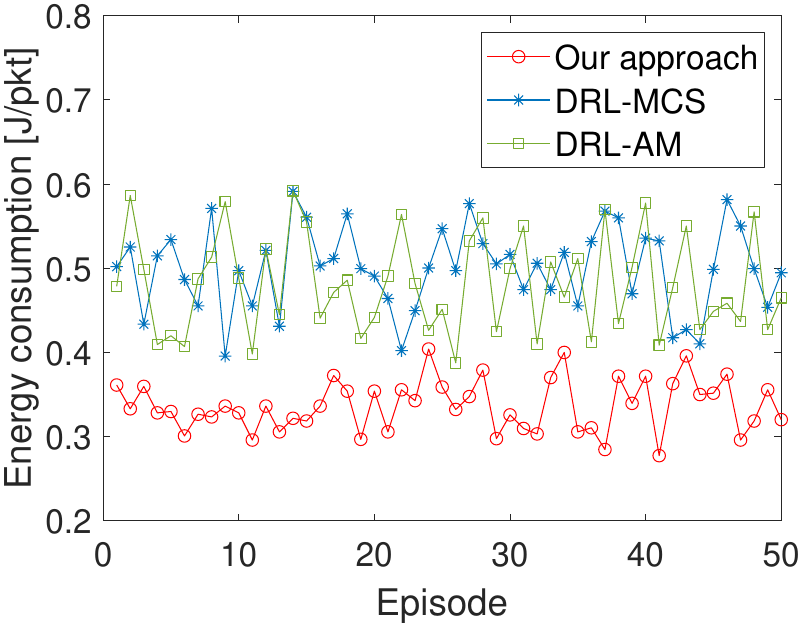}
    \caption{10-node scenario.}
    \label{fig:10nodesener}
\end{subfigure}

\caption{Energy consumption comparison between our approach, DRL-MCS, and DRL-AM.}
\label{fig:energy}
\end{figure*}

The interval is selected using a standard UCB strategy.

\begin{equation} Q_k = \arg\max_{Q \in \mathcal{Q}} \left( \hat{\mu}_k(Q) + \sqrt{\frac{c\log k}{N_k(Q)}} \right), \end{equation}
where $\hat{\mu}_k(Q)$ is the empirical average reward for action $Q$, $c$ is the exploration–exploitation parameter, $k$ is the current round index, and $N_k(Q)$ is the number of times $Q$ has been selected.

\section{Simulation scenario} \label{sim}

To evaluate performance, we ran simulations with 4, 6, 8, and 10 UW nodes. Scalability is generally not a major concern in UWA networks, which rely on few nodes with long transmission ranges. The high deployment cost also limits large-scale setups. The topology ensured a maximum node distance of 357 meters, similar to the LOON testbed \cite{alves2014loon}. 

Simulation parameters reflect real-world UW devices, such as EvoLogics modems \cite{evologics}, using a 10.5 kHz carrier and 4.2 kHz bandwidth. Nodes transmitted 125-kilobyte data packets and control packets at 4800 bps. These settings capture UWA-specific challenges, including multipath, Doppler spread, and noise variability. The environment simulated wind speed $w = 50$ km/h, shipping factor $z = 0.5$, and spreading factor $s = 1.75$, introducing link quality variations from obstacles and wave motion. The framework continuously adapts to such dynamics. \textcolor{black}{The learning process is organized into \textit{episodes}, each consisting of a sequence of \textit{decision epochs} where the agent selects an action. The episode performance corresponds to the average performance over all its decision epochs.}

To assess the proposed \textcolor{black}{bilevel} MAB framework, which jointly selects \textcolor{black}{Modulation Scheme (MS), transmission power level $P$, and feedback interval $Q_k$}, we compare it with two \textcolor{black}{existing DRL} solutions for \textcolor{black}{adaptive modulation}: the Deep Reinforcement Learning-based Adaptive Modulation (DRL-AM) algorithm \cite{mashhadi2021deep} and the DRL-based intelligent Modulation and Coding Scheme selection (DRL-MCS) algorithm \cite{8703432}. Note that, due to the lack of efficient RL solutions specifically designed for UW networks, we have considered the two aforementioned algorithms as they proved to be effective in challenging terrestrial networks characterized by the coexistence of multiple users and/or limited Channel State Information (CSI). Specifically:

\begin{itemize} 
\item \textit{DRL-AM} addresses adaptive modulation under outdated CSI using a DRL framework to dynamically adjust MSs. Its robustness to non-linear channel variations makes it relevant for UW environments.
\item \textit{DRL-MCS} targets cognitive networks with primary and secondary users sharing the same spectrum. It optimizes modulation and coding to minimize interference from secondary users and improve overall performance.
\end{itemize}

\begin{figure} [t]
    \centering
    \includegraphics[width=0.74\linewidth]{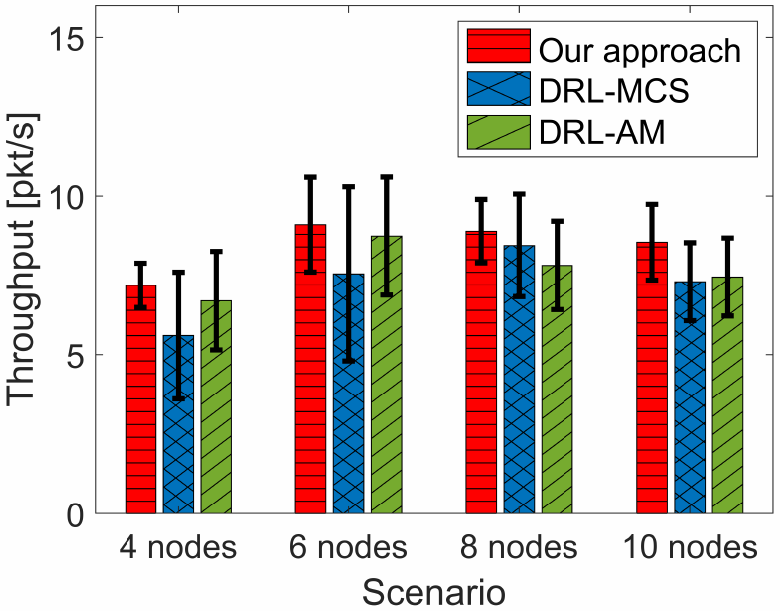}
    \caption{Throughput in different scenarios.}
    \label{fig:hystosim}
\vspace{-0.3cm}
\end{figure}

As both DRL-based solutions are computationally intensive, we assume they operate at the Sink node rather than on resource-constrained IoUT devices. They adopt a centralized approach, applying the same MS to all nodes without per-node optimization. In addition, they do not adjust $P$ nor control the feedback interval. As a result, each action requires immediate feedback, increasing network overhead. Simulations ran for a total duration of 6000 seconds. Both the \textcolor{black}{bilevel} MAB framework and the DRL baselines were trained on a high-performance system with an Nvidia GeForce RTX 4090 GPU. For MAB agents, $c$ was set to $2$, and $\theta=0.7$.

\section{Performance evaluation} \label{num-results}

This section analyzes the performance of our framework, comparing it against the DRL-based solutions under different network configurations.

\subsection{Throughput Analysis}

We evaluate our framework by analyzing training throughput across 4- to 10-node scenarios, as shown in Fig.~\ref{fig:throughput}, which details the 8- and 10-node cases. Our approach consistently outperforms DRL-based baselines in all scenarios, \textbf{achieving up to 20.61\% higher throughput than DRL-MCS and 14.75\% over DRL-AM}. This improvement is due to its ability to adapt MS, $P$, and feedback interval to local channel conditions in a fully distributed and low-variance fashion.

Unlike DRL-based solutions, which suffer from oscillatory behavior and centralized bottlenecks, our framework converges more rapidly and stably across different network sizes. DRL-AM and DRL-MCS apply uniform settings across nodes and rely on global coordination, limiting their adaptability in dense and heterogeneous environments.

Fig.~\ref{fig:hystosim} confirms this trend, showing that packet loss grows with network density. Our distributed scheme mitigates the issue by letting each node optimize transmissions via local feedback, ensuring high throughput and consistent performance even as the network scales.

\begin{figure}
    \centering
    \includegraphics[width=0.74\linewidth]{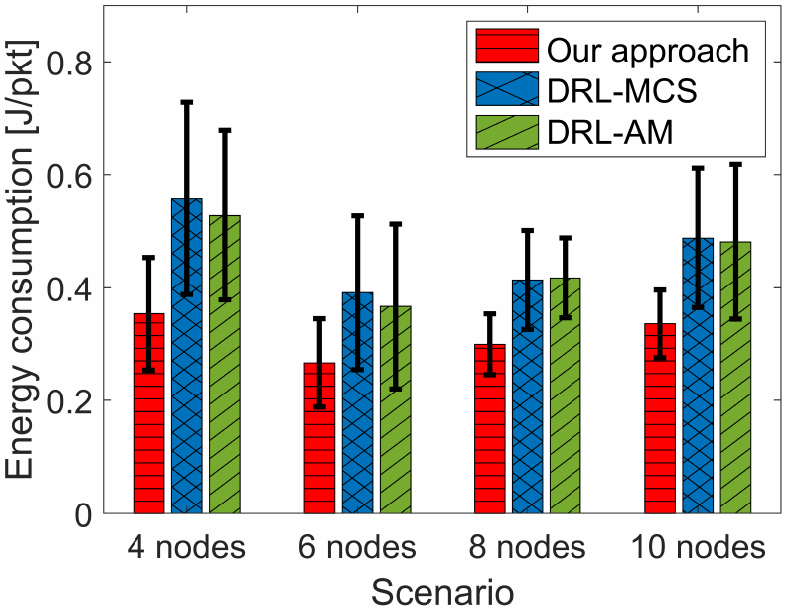}
    \caption{Energy efficiency in each considered scenario.}
    \label{fig:energy_consumption}
\end{figure}

\begin{figure} [t]
    \centering
    \includegraphics[width=0.74\columnwidth]{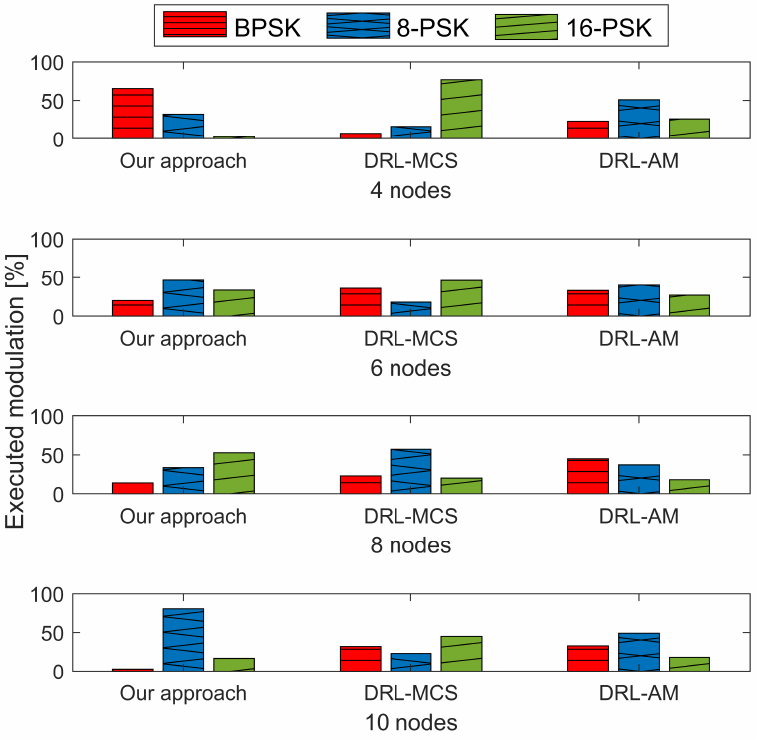}
    \caption{Modulation selection frequency for our approach, DRL-MCS and DRL-AM in each considered scenario.}
    \label{fig:modulations}
\end{figure}

\subsection{Energy Consumption Analysis}

We further assess our framework by analyzing energy consumption across all the network scenarios, as shown in Fig.~\ref{fig:energy}, highlighting the 8- and 10-node cases. Our framework systematically surpasses DRL baselines, \textbf{achieving up to 36.60\% lower energy consumption than DRL-MCS and 33.05\% compared to DRL-AM}. This improvement results from the joint optimization of transmission settings and feedback frequency.
Two factors drive this gain. First, by jointly selecting MS and $P$, each node transmits using only the energy needed for successful delivery. Second, by dynamically adjusting the feedback interval $Q_k$, the framework reduces control transmissions when frequent updates are unnecessary.

In contrast, \textcolor{black}{prior} DRL-based solutions lack both power control and feedback scheduling. Feedback is sent after every packet and $P$ remains fixed, leading to excessive signaling and higher energy consumption. Fig.~\ref{fig:energy_consumption} confirms this behavior. By adapting $P$ and $Q_k$ to channel dynamics and throughput needs, the framework lowers both transmission and feedback energy, preserving performance.

\subsection{Adaptive Modulation Strategy}

Fig.~\ref{fig:modulations} shows the modulation selection frequency across different scenarios for our framework and the DRL baselines.

\begin{table}[]
\centering
\begin{tabular} {|c|c|c|c|}
\hline
Scenario & 4 min & 7 min & 10 min \\\hline
4 nodes  & 48.40\% & 31.41\% & 20.19\% \\\hline
6 nodes  & 61.78\% & 28.11\% & 10.11\% \\\hline
8 nodes  & 57.83\% & 30.50\% & 11.67\% \\\hline
10 nodes & 50.80\% & 33.93\% & 15.27\% \\\hline
\end{tabular}
\caption{Average distribution of feedback interval selections $Q_k$ across different network scenarios.}
\label{tab:feedback_distribution}
\end{table}

Our approach dynamically selects the MS at each node by accounting for both network density and local channel conditions. In sparse scenarios (e.g., 4 nodes), lower-order modulations like BPSK are preferred for their robustness to channel variability. As density increases (e.g., 6–8 nodes), the agent shifts toward higher-order schemes such as 8-PSK and 16-PSK to exploit better connectivity and boost throughput.

In high-density cases (e.g., 10 nodes), higher-order modulations remain common, but with reduced variation. This reflects the need to balance spectral efficiency with resilience to interference in crowded acoustic environments.

\textcolor{black}{In contrast, DRL-based schemes enforce uniform MS across nodes, ignoring channel diversity---hurting strong links and degrading weaker ones. This limited adaptability reduces efficiency and responsiveness, explaining the suboptimal use of 16-PSK even in sparse scenarios.}

\subsection{Feedback Interval Analysis}

Table~\ref{tab:feedback_distribution} reports the average distribution of feedback interval selections $Q_k$ across different network sizes. As node density increases, the agent progressively favors shorter intervals to maintain context freshness.

In the 4-node case, the distribution is balanced, with 4-minute updates being most frequent (48.40\%) but not dominant. The presence of 10-minute intervals (20.19\%) suggests occasional feedback reduction under stable conditions. With 6 and 8 nodes, the preference for 4-minute intervals strengthens (61.78\% and 57.83\%), reflecting the need for more frequent updates. The persistent use of 7-minute intervals (28.11\% and 30.50\%) indicates a trade-off between feedback cost and freshness. In the 10-node case, the share of 10-minute intervals rises again (15.27\%), likely to mitigate signaling overhead. The slight drop in 4-minute updates (50.80\%) suggests that the agent relaxes feedback frequency as network load increases.

Overall, the agent adjusts $Q_k$ to balance throughput and signaling cost, rather than aiming to minimize AoI directly.

\section{Conclusion} \label{conclusion}

In this paper, we proposed a fully distributed \textcolor{black}{bilevel} MAB framework for UnderWater Acoustic (UWA) networks that jointly optimizes Modulation Schemes (MSs), transmission power ($P$), and feedback intervals. The first level uses a Contextual Delayed MAB (CD-MAB) to adapt transmission decisions from delayed feedback, while the second level leverages a Feedback Scheduling MAB to dynamically regulate the feedback interval, balancing throughput and signaling overhead. The fully distributed design ensures scalability and low overhead, suiting Internet of Underwater Things (IoUT) scenarios and enabling future extensions to non-stationary settings and physical-layer models. Simulations with the DESERT UW simulator show our framework outperforms DRL baselines, \textbf{achieving up to 20.61\% higher throughput than DRL-MCS and 14.75\% over DRL-AM, while cutting energy use by up to 36.60\% and 33.05\%, respectively, with faster convergence and improved stability}.

\section*{Acknowledgment}

The authors thank Sirin Chakraborty from Auburn University for helpful discussion on AoI.

\bibliographystyle{IEEEtran}
\bibliography{bibliography.bib}

\end{document}